\documentclass{emulateapj}

\usepackage{apjfonts}
\usepackage{amssymb}
\usepackage{epsfig}
\usepackage{amsmath}
\usepackage{graphicx} 
\usepackage{relsize}
\usepackage[toc,page]{appendix}
\usepackage[T1]{fontenc}
\usepackage{chngcntr}

\usepackage[backref,breaklinks,colorlinks,citecolor=blue]{hyperref}
\usepackage[all]{hypcap}

\usepackage[caption = false]{subfig}

\newcommand{\kms}{\hbox{km~s$^{-1}$}}

\newcommand{\ciii}{\hbox{{\ion{C}{3}}]}}
\newcommand{\siiii}{\hbox{{\ion{Si}{3}}]}}

\begin{document}

 \title{The Complete Ultraviolet Spectrum of the Archetypal
   ``Wind-Dominated'' Quasar Mrk~231: Absorption and Emission from a
   High-Speed Dusty Nuclear Outflow}

\author{S. Veilleux\altaffilmark{1,2},
  M. Mel\'endez\altaffilmark{1,3,4}, T. M. Tripp\altaffilmark{5},
  F. Hamann\altaffilmark{6,7}, \& D. S. N. Rupke\altaffilmark{8}}

\altaffiltext{1}{Department of Astronomy, University of Maryland,
  College Park, MD 20742, USA; veilleux@astro.umd.edu}

\altaffiltext{2}{Joint Space-Science Institute, University of Maryland,
  College Park, MD 20742, USA}

\altaffiltext{3}{NASA Goddard Space Flight Center, Greenbelt, MD
  20771, USA}

\altaffiltext{4}{Wyle Science, Technology and Engineering Group, 1290
  Hercules Avenue, Houston, TX 77058 USA}

\altaffiltext{5}{Department of Astronomy, University of Massachussetts,
  Amherst, MA 01003, USA}

\altaffiltext{6}{Department of Physics and Astronomy, University of
  California, Riverside, CA 92507, USA}

\altaffiltext{7}{Department of Physics \& Astronomy, University of
  California, Riverside, CA 92521, USA}

\altaffiltext{8}{Department of Physics, Rhodes College, Memphis, TN
  38112, USA}

\begin{abstract}
New near- and far-ultraviolet (NUV and FUV) {\em HST} spectra of
Mrk~231, the nearest quasar known, are combined with ground-based
optical spectra to study the remarkable dichotomy between the FUV and
NUV-optical spectral regions in this object. The FUV emission-line
features are faint, broad, and highly blueshifted (up to $\sim 7000$
km s$^{-1}$), with no significant accompanying absorption. In
contrast, the profiles of the NUV absorption features resemble those
of the optical Na~I~D, He~I, and Ca~II H and K lines, exhibiting broad
blueshifted troughs that overlap in velocity space with the FUV
emission-line features and indicate a dusty, high-density and patchy
broad absorption line (BAL) screen covering $\sim$90\% of the observed
continuum source at a distance $\la$2 -- 20 pc. The FUV continuum
emission does not show the presence of any obvious stellar features
and is remarkably flat compared with the steeply declining NUV
continuum.  The NUV (FUV) features and continuum emission have not
varied significantly over the past $\sim$22 (3) years and are
unresolved on scales $\sim$40 (170) pc.  These results favor an AGN
origin for the NUV -- FUV line and continuum emission.  The observed
FUV line emission is produced in the outflowing BAL cloud system,
while the Balmer lines arise primarily from the standard broad line
region seen through the dusty BAL screen.  Our data are inconsistent
with the recently proposed binary black hole model. We argue instead
that Mrk~231 is the nearest example of weak-lined ``wind-dominated''
quasars with high Eddington ratios and geometrically thick (``slim'')
accretion disks; these quasars are likely more common in the early
universe.
\end{abstract}


\section{Introduction}

Observations and numerical simulations have shown that galactic winds
play a crucial role in the story of galaxy evolution (see Veilleux,
Cecil, \& Bland-Hawthorn 2005; Fabian 2012 for reviews). They are the
primary mechanism by which matter, energy, and metals are recycled in
galaxies and are deposited into the surrounding circumgalactic,
intracluster, or intergalactic medium (CGM, ICM, or IGM, respectively;
e.g., Tumlinson et al.\ 2011; Tripp et al.\ 2011; Werk et al.\ 2013,
2014; Rubin et al.\ 2014, 2015; Prochaska et al.\ 2014; Lau et
al.\ 2015, and references therein).  Galactic winds may also quench
star formation by heating the cool gas and/or ejecting it from the
host.  For instance, only a few percent of the radiative energy
liberated by the growth of a supermassive black hole (SMBH) needs to
be tapped in this way to truncate star formation and the growth of the
stellar mass. These ``quasar-mode'' outflows are distinct from
``radio-mode'' jets in that they are much less collimated, and
therefore have the potential to impact a much greater swath of a
galaxy's gas.  Quasar-mode outflows are often invoked to play a
fundamental role in the evolution of both SMBHs and their host
galaxies, quenching star formation and regulating black hole
growth. They may be at the origin of the tight SMBH -- galaxy
relations (e.g., Murray, Quataert, \& Thompson 2005; Di Matteo,
Springel, \& Hernquist 2005) and possibly also the cutoff of the
galaxy mass function at the high stellar mass end (e.g., Croton et
al.\ 2006; Hopkins et al.\ 2008).

Direct observational evidence for large-scale quasar mode feedback in
galaxies, near and far, has grown dramatically in recent years
(molecular gas phase: e.g., Veilleux et al.\ 2013b; Spoon et
al.\ 2013; Cicone et al.\ 2014; Gonzalez-Alfonso et al.\ 2016; neutral
and ionized gas phases: e.g., Maiolino et al.\ 2012; Rupke \& Veilleux
2013; Harrison et al.\ 2014; Zakamska \& Greene 2014; Cicone et
al.\ 2015; Cresci et al. 2015; Brusa et al. 2016; Zakamska et
al.\ 2015, 2016).  Mrk~231, the focus of the present paper, is the
nearest quasar known ($z$ = 0.0422, corresponding to a
distance\footnote{For a cosmology with $H_0$ = 73 km s$^{-1}$
  Mpc$^{-1}$, $\Omega_{\rm matter}$ = 0.27, and $\Omega_{\rm vacuum}$
  = 0.73.} of 178 Mpc, or 1$\arcsec$ = 0.863 kpc) and arguably the
best laboratory to study this quasar feedback in action. It is
experiencing a powerful multi-phase and multi-scale outflow that has
the potential to rapidly transform the host gas-rich galaxy merger
remnant into a ``red and dead'' spheroidal galaxy (e.g, Rupke et
al.\ 2005; Fischer et al.\ 2010; Sturm et al.\ 2011; Rupke \& Veilleux
2011, 2013; Veilleux et al.\ 2014; Aalto et al.\ 2012; Cicone et
al.\ 2012; Feruglio et al.\ 2010; 2015; Alatalo 2015; Lindberg et
al.\ 2016, and references therein).

Mrk~231 was the first target observed under our Cycle 19
far-ultraviolet (FUV) spectroscopic survey of the nearby {\em QUEST}
({\em Q}uasar -- {\em U}LIRG {\em E}volution {\em St}udy; e.g.,
Veilleux 2012 and references therein) QSOs with the Cosmic Origins
Spectrometer (COS) on board the {\em Hubble Space Telescope} ({\em
  HST}).  The unprecedented sensitivity of COS allowed us to obtain
the first high-S/N spectrum of Mrk~231 below $\sim$1500 \AA\ (Veilleux
et al.\ 2013a; hereafter Paper I). This spectrum is highly peculiar,
highlighted by the presence of faint ($\la$2\% of predictions based on
H$\alpha$), broad ($\ga$ 10,000 km s$^{-1}$ at the base), and highly
blueshifted (centroid at $\sim$ --3000 km s$^{-1}$) Ly$\alpha$
emission (all velocities reported in this paper are in the reference
frame of the quasar, where $v$ = 0 km s$^{-1}$ at redshift $z = 0.0422$,
derived from CO and H~I observations; Downes \& Solomon 1998; Carilli
et al.\ 1998).  The FUV continuum emission does not show the presence
of any obvious photospheric or wind stellar features (e.g., James et
al.\ 2014) and is remarkably flat compared with the steeply declining
NUV continuum (Smith et al.\ 1995; Paper I; Leighly et al.\ 2014).

Surprisingly, the FUV spectrum also does not show any unambiguous
broad absorption features. This is unexpected because the core of
Mrk~231 harbors well-known spatially unresolved broad absorption-line
(BAL) systems in the optical and near-ultraviolet bands (Boksenberg et
al.\ 1977; Smith et al.\ 1995; Rupke, Veilleux, \& Sanders 2002;
Gallagher et al.\ 2002, 2005). The Mrk~231 BAL features are detected
in several transitions including many low-ionization species (e.g., Na
I D $\lambda\lambda$5890, 5896, He I$^*$ $\lambda$3889, Ca II H and K,
Mg~II $\lambda\lambda$2796, 2803, and Mg~I $\lambda$2853, as
well as Fe~II UV1 and Fe~II UV2, hence fitting the rare FeLoBAL
category). They show strong absorption across the range $\sim$ --3500
to --5500 km s$^{-1}$ and no significant absorption at velocities
above --3500 km s$^{-1}$ (e.g., Rupke et al.\ 2002; Paper I).  The
absence of narrow ISM lines or stellar P-Cygni profiles in the COS
spectrum suggests that the FUV continuum is dominated by AGN
emission. However, the BAL features, which are usually attributed to
AGN-driven outflows, are also not evident in the FUV spectrum (Paper
I). This suggests that the BAL outflow is highly dusty but is also
patchy so that the FUV flux can leak through the FeLoBAL screen
without having BAL features imprinted on the spectrum.  The conclusion
that dust plays virtually no role in the FUV is supported by the
flatness of the FUV continuum and the near-zero polarization at these
wavelengths measured from spectropolarimetric data with the Faint
Object Spectrograph (FOS; Smith et al.\ 1995).  The observed
Ly$\alpha$ emission is best explained if it is produced in the
outflowing BAL cloud system, while the Balmer lines arise primarily
from the standard broad emission line region (BELR) seen through the
dusty ($A_V \sim 7$ mag.\footnote{Leighly et al.\ (2014) have argued
  that the amount of extinction may be as low as $A_V \sim 1.5 - 1.6$
  mag. when applying the reddening curve of Goodbar (2008), previously
  used to explain low values of total-to-selective extinctions in Type
  Ia supernovae.})  broad absorption line region (BALR).

Many of these results were explained in Paper I using a simple
spherical geometry or a physically motivated but slightly more complex
disk geometry (Figures 5 and 6 in Paper I). However, it was noted that
the lack of a contemporaneous high-S/N spectrum covering the
diagnostic-rich region from 1500 \AA\ to 3000 \AA\ was a major
hindrance to further interpretation (a 1992 pre-COSTAR FOS NUV spectrum at
2500 -- 3300 \AA\ published by Smith et al.\ 1995 was used in
the analysis of Paper I). We recently addressed this problem with the
acquisition in 2014 of high-S/N FUV and NUV spectra with {\em HST} COS
and STIS, respectively, as well as near-contemporaneous ground-based
optical spectra with the 4.3-meter Discovery Channel Telescope
(DCT). This paper reports the results of our analysis of these new
spectra and comparisons with the data used in Paper I.  The
acquisition and reduction of the {\em HST} and DCT spectra are
discussed in Section 2. The results derived from the new data and
comparisons with the old data are described in Section 3. The
geometrical models proposed in Paper I are revisited in Section 4,
using the additional constraints provided by the new data as well as
recent results in the literature. The results are also compared
against the predictions from the binary AGN model recently proposed to
explain the ultraviolet spectrum of this object by Yan et al.\ (2015).

\section{Observations}

\subsection{{\em HST} Ultraviolet Spectroscopy}

Mrk~231 was revisited using the G140L grating on COS (6 orbits on 2014
December 13) and the G230L grating on STIS (1 orbit on 2014 July 1).
The first 5 orbits of the COS/G140L observations were centered at 1280
A, while the other COS orbit was centered at 1105 \AA\ to cover the
redshifted Ly$\alpha$ region (which is in the chip gap at 1280
\AA). Overall, the COS spectra cover $\sim$1100 -- 1900
\AA\ ($\sim$1050 -- 1825 \AA\ in the rest frame) at a spectral
resolving power of $R = (\lambda/\delta\lambda) \sim 2000 - 5000$. The
STIS spectrum covers the NUV range from 1800 -- 3180 \AA\ (1750 --
3050 \AA\ in the rest frame) at a spectral resolving power of $R \sim
500$, which is sufficient to characterize the BEL and BAL systems in
the NUV lines. Note that the Cycle 19 COS spectrum from Paper I was
obtained with G130M and therefore had a higher spectral resolution but
more limited spectral coverage ($\sim$1150 -- 1470 \AA) than the new
COS spectrum.

The COS observations used standard target acquisition procedures.  For
STIS, we used F28X50OII to center on the bluest part of the spectrum
to minimize possible (but unlikely for this QSO) centroid shifts
between the NUV and optical positions. All of the COS observations
were done in TIME-TAG mode.  The target exposures were split into four
segments using focal-plane splits. This observing strategy was adopted
to reduce the fixed pattern noise. The total exposure times were
12,300 seconds at 1280 \AA\ and 2660 seconds at 1105 \AA.  The STIS
exposures were also split into four segments of equal length (473
seconds to take maximum use of the visility), for a total exposure
time of 1892 seconds.

All observations were processed using both the standard {\em
  calcos}\footnote{See Chapter 3 of Part II of the HST Data Handbook
  for COS
  (http://www.stsci.edu/hst/cos/documents/handbooks/datahandbook) for
  the details of the calcos processing.} calibration pipeline and a
custom-made pipeline put together by one of us (T.M.T.) to optimize
the quality of the resulting data. The results from both pipelines
were found to be nearly identical (the data presented in this paper
are from the custom-made pipeline).

\subsection{Ground-based Optical Spectroscopy}

To minimize the effects of spectral variability (e.g., the
highest-velocity component of the BAL systems in Mrk~231 at $\sim$
--8000 km s$^{-1}$ is known to be time-variable; Hutchings \& Neff
1987; Boroson et al.\ 1991; Kollatschny et al.\ 1992; Forster et
al.\ 1995; Rupke et al.\ 2002, 2005), we acquired a new optical
spectrum of this object to complement our 2001, 2004, 2007, and 2012
spectra published in Rupke et al.\ (2002), Rupke et al.\ (2005c),
Rupke \& Veilleux (2011), and Paper I, respectively. The deVeny
spectrograph on the 4.3-meter DCT was used on 2015 June 8 and 9 with
the 300 g mm$^{-1}$ grating in the first order and a 1$\farcs$5 slit
to obtain 8 spectra each of 5 min duration with a spectral resolving
power of $\sim$920, covering $\sim$3530 -- 8000 \AA. Standard
IRAF\footnote{http://iraf.noao.edu/} procedures were used to reduce
and calibrate these spectra. The sky conditions were not photometric
and the flux calibration below 5500 \AA\ was found to be unreliable so
this portion of the DCT spectrum is not discussed here.

\section{Results}

The 2014 Cycle 21 HST/COS and STIS spectra of Mrk~231 are shown in
Figures 1-4 along with the 2011 October 15 Cycle 19 {\em HST}/COS
spectrum, the 2015 DCT/deVeny optical spectrum, and the 2001 Keck/ESI
optical spectrum for comparison. The line measurements derived from
these data are listed in Table 1.

Figure 1 shows that the 2011 and 2014 COS spectra are remarkably
similar: there is no evidence for any variations in the FUV continuum
level and slope, within the uncertainties of the measurements. At both
epochs, the FUV continuum from $\sim$1100 -- 1700 \AA\ (rest frame)
declines slightly at shorter wavelengths, following $F_\lambda \propto
\lambda^{\alpha_\lambda}$ where $\alpha_\lambda$ = 0.7 $\pm$ 0.3 using
anchor regions at [1298 -- 1304], [1416 -- 1418] and [1508 -- 1515]
\AA\ ($\alpha_\lambda$ increases to $\sim$1.7 when considering only
the continuum below Ly$\alpha$; see Paper I).  Similarly, we find no
evidence for change in the Ly$\alpha$ profile between 2011 and 2014,
within the uncertainties of our measurements (dominated by
uncertainties in the continuum placement and difference in spectral
resolution between the two observations).

The profile of the Ly$\alpha$ emission, shown in detail in Figure 5,
has remained highly peculiar, showing a narrow peak centered at
$\sim$1261 \AA\ ($\sim$ --1500 km s$^{-1}$ in the rest frame) and
wings extending over at least 1240 -- 1265 \AA\ ($\sim$6000 km
s$^{-1}$) and quite possibly as much as 1225 -- 1280
\AA\ ($\sim$13,000 km s$^{-1}$). The overall centroid of the
Ly$\alpha$ emission is blueshifted by $\sim$ 3000 km s$^{-1}$ with
respect to rest (given the absence of Si~II $\lambda$1260 and O~I
$\lambda$1302 emission, Si~II $\lambda\lambda$1190, 1193, N~I
$\lambda$1200, and Si~III $\lambda$1207 emission do not affect the
profile of Ly$\alpha$). The broad width and large blueshift clearly
point to an AGN origin for this feature. This is made clear in Figure
5, where we compare the profile of Ly$\alpha$ with those of H$\alpha$
emission and Na I D $\lambda\lambda$5890, 5896 absorption, derived
from the most recent DCT data. Since we find no evidence for
significant changes in the profiles of H$\alpha$ and Na I~D since 2001
(Rupke et al.\ 2002), we also show in Figure 5 the profiles of He I$^*
\lambda$3889 and Ca II H and K derived from these older data 
  (Na~I~D is distinctly broader than these features because it is a
  blended doublet transition). The high velocities of these features
are consistent with material originating close to the central black
hole.  Indeed, the Ly$\alpha$ emission is centered on the NUV nucleus
and spatially unresolved with the rather coarse spatial resolution of
COS at 1250 \AA\ for our settings ($\sim$1$\farcs$3 $\sim$ 1.1 kpc;
Ghavamian et al.\ 2010). Similarly, the broad H$\alpha$ emission and
highly blueshifted Na I D troughs are spatially unresolved
($\la$0$\farcs$6 $\sim$ 500 pc) in the ground-based IFU data of Rupke
\& Veilleux (2011, 2013).  Better constraints on the size of the
source of FUV-NUV emission are derived from the excellent match in the
overlap region between the new COS and STIS spectra (see below).

As noted in Paper I, the Ly$\alpha$ emission is also much fainter
relative to H$\alpha$ than the prediction from Case B recombination
(Ly$\alpha$/H$\alpha$ $\sim$ 0.05 instead of the expected
$\sim$13). While collisional suppression of the level 2 population in
hydrogen is known to reduce Ly$\alpha$/H$\alpha$ ratio by a factor of
up to $\sim$2 -- 3 relative to the Case B value in the high-density
BELRs of quasars (e.g., Netzer et al.\ 1995), the more extreme
Ly$\alpha$/H$\alpha$ ratio of Mrk~231 is likely due to Ly$\alpha$
absorption by dust grains. This conclusion seems inescapable when
considering the gas at $v \ge 0$ \kms\ where Ly$\alpha$/H$\alpha$ is
even more extreme. This issue is revisited in Section 4.

The new COS data confirm the dearth of clearly identifiable features,
other than Ly$\alpha$, within the wavelength coverage of the 2011 COS
spectrum (see Table 1 in Paper I for a list of tentative
identifications for the various ``bumps'' and ``wiggles'' seen in the
2011 spectrum and also seen in the recent spectrum).  However, the new
COS/G140L spectrum now extends to $\sim$1900 \AA\ and shows a
significant detection of the C~IV $\lambda\lambda$1548, 1550 emission
feature (Figure 1). This feature appears purely in emission, and is
broad ($\sim$6000 km s$^{-1}$) and highly blueshifted ($\sim$3250 km
s$^{-1}$), i.e. similar to the Ly$\alpha$ emission profile (see
comparison in Fig.\ 6). The C~IV absorption reported by Gallagher et
al.\ (2002) in the 1992 FOS data is not present in the newer 2014
data. The detected C~IV emission is faint, representing only
$\sim$20\% of the (faint) Ly$\alpha$ emission (Table 1).

The new STIS spectrum presented in Figure 2 also shows the presence of
faint, highly blueshifted \ciii\ $\lambda$1908.7 line emission with a
profile that resembles that of Ly$\alpha$ (Fig.\ 6). As first pointed
out by Smith et al.\ (1995) and Gallagher et al.\ (2002), and shown in
Figure 6, one has to be careful when interpreting this emission
feature since there are several other transitions in this wavelength
region that may potentially contaminate the \ciii\ profile. The
absence of Fe III UV 34 $\lambda$1926.3 suggests that the Fe III UV 34
multiplet transitions at 1895.5 and 1914.1 \AA\ do not significantly
affect the profile of \ciii. Similarly, \siiii\ $\lambda$1892.0 is
unlikely to be important as it is generally undetected or much fainter
than \ciii\ and Fe~III in BAL quasars (e.g,. Hartig \& Baldwin 1986;
Hall et al.\ 2002; although it can be strong in non-BAL quasars: e.g.,
Laor et al.\ 1997 in I~Zw~1 and Baldwin et al.\ 1996 in a sample of
narrow-line quasars).  Finally, Al III $\lambda\lambda$1854.7, 1862.8
are too far to the blue to significantly affect the bulk of the
\ciii\ profile (only perhaps the extreme blue wing).

For the sake of completenes, we also show in Figure 6 the
continuum-subtracted emission near the expected location of C~II
$\lambda$1334. As pointed out in Paper I, the S/N ratio here is low
and continuum placement is tricky, making this detection only
tentative. However, we note that the profile of this faint emission is
similar within the uncertainties to that of the other FUV lines,
providing tantalizing evidence for blueshifted C~II $\lambda$1334
emission in this spectrum.

As first noted by Smith et al.\ (1995), and discussed in detail in
Paper I and Leighly et al.\ (2014), a dramatic upturn in the spectral
energy distribution (SED) is seen above $\sim$2200 \AA\ (Fig.\ 4).
Remarkably, the spectrum longward of this wavelength is also rich in
deep blueshifted absorption throughs, in stark contrast with the few
faint and blueshifted emission lines detected in the FUV. The main NUV
features detected in the STIS spectrum are the broad absorptions lines
from He~I $\lambda$2945.1, Mg I $\lambda$2852.9, Mg II
$\lambda\lambda$2796.3, 2803.4, and the Fe II UV1 UV2, UV3, UV61,
UV62, UV63, and UV78 multiplets.  The strongest of these features are
labeled in Figures 2 and 3 and shown in velocity space in Figure
7. The purpose of this last figure is not to compare the profiles
point-by-point, since these features range in complexity from 
  single transitions (He~I and Mg~I), blended doublet (Mg~II), and
multiple transitions from various ground and excited states making a
detailed comparison meaningless, but to emphasize that all of features
are produced in the outflowing material.  These features are common in
the rest-frame UV spectra of FeLoBAL quasars (e.g., Hall et al.\ 2002;
Lucy et al.\ 2014), and all of them except Fe II UV3, UV60, UV61, and
UV78 were identified in the 1992 FOS spectrum first presented by Smith
et al.\ (1995). The presence of broad absorption features from various
excited states provides important constraints on the properties of the
BALR. We return to this point at the end of Section 4.2.

A direct comparison between the 2014 STIS NUV spectrum and the 1992
FOS spectra obtained by Smith et al.\ (1995) is made difficult by the
limited S/N and significant ($\ge$30\%) light losses of the aberrated
pre-COSTAR FOS data. Figure 2 shows a comparison after heavily
smoothing the FOS data and scaling them to try to account for these
light losses and reproduce the spectra shown in Figure 1 of Smith et
al.  Interestingly, the 2014 STIS NUV spectrum obtained through a
0$\farcs$2 $\times$ 50\arcsec\ aperture is remarkably similar to the
1992 FOS spectra obtained through an effective aperture of 4$\farcs$3
$\times$ 1$\farcs$4. The limited S/N of the FOS data does not allow us
to carry out a detailed analysis of the various features in these
data, but a visual comparison of the two spectra does not reveal any
significant differences in the various absorption features. This
indicates that the NUV continuum emission and BAL features have
remained constant over a period of at least $\sim$22 years and are
spatially unresolved on a scale of $\la$0$\farcs$2 $\sim$ 170 pc.  The
NUV target acquisition images obtained in both Cycles 19 and 21 are
consistent with an unresolved point source centered on the optical
position of the quasar at the resolution of the COS imager (intrinsic
FWHM $\la$ 2 pixels $\sim$ 0$\farcs$05 $\sim$ 40 pc).  Similarly, all
of the NUV continuum and line emission is unresolved along the slit in
the STIS spectrum (intrinsic FWHM $\la$ 1 pixel $\sim$ 0$\farcs$05
$\sim$ 40 pc). Finally, we note that the flux level of the 2014 STIS
NUV spectrum obtained with the 0$\farcs$2 $\times$ 50\arcsec\ aperture
is consistent with that of the 2014 COS FUV spectrum obtained with a
2$\farcs$5 $\times$ 2$\farcs$5 aperture in the overlap region. Barring
fortuitious counteracting effects, this suggests that not only is the
NUV emission unresolved on scales below $\sim$170 pc scale but also
the FUV emission. This is another argument in favor of an AGN origin
for both the NUV and FUV continuum emission.

\section{Discussion}

The new FUV, NUV, and optical spectra presented in this paper add
strong new constraints on the structure and geometry of the BELR,
BALR, and source of continuum emission in Mrk~231. First, in Section
4.1, we compare the UV line and continuum properties of Mrk~231 with
those of other quasars, drawing an analogy with PHL-1811 analogs and
weak-lined quasars (WLQs).  Next, in Section 4.2, we revisit the
geometrical models presented in Paper I to interpret the new
data. Finally, in Section 4.3, we confront the predictions from the
binary AGN scenario proposed by Yan et al.\ (2015) with the results of
our analysis of the new data.

\subsection{Far-Ultraviolet Properties of Mrk~231 in the Broader Context}

As recently pointed out by Teng et al.\ (2014), Mrk~231 is an outlier
among quasars, in the sense that it is weaker in the X-rays than
typical quasars: the total absorption-corrected AGN luminosity for
Mrk~231 in 0.5 -- 30 keV is only $\sim$1.0 $\times$ 10$^{43}$ erg
s$^{-1}$. Compared to L$_{\rm bol,AGN}$ of 1.1 $\times$ 10$^{46}$ erg
s$^{-1}$ (Veilleux et al.\ 2009), the 2 -- 10 keV X-ray luminosity is
only 0.03-0.05\% of the AGN bolometric luminosity. This ratio is
typically $\sim$2 to 15\% among radio-quiet quasars, with the most
luminous objects typically having the lowest ratios. However, X-ray
weakness is common among Seyfert 1-like ULIRGs (Teng \& Veilleux
2010), LoBAL QSOs (Luo et al.\ 2013), PHL-1811 analogs (Leighly et
al.\ 2007a,b; Wu et al.\ 2011), and other ``wind dominated''
weak-lined quasars (WLQs; Luo et al.\ 2015). Evidence for significant
X-ray absorption is present in many of these objects, particularly the
PHL-1811 analogs and WLQs (Luo et al.\ 2015), but not in PHL~1811
itself (Leighly et al.\ 2007a), nor Mrk~231 (Teng et al.\ 2014).

Interestingly, the accretion rate in many of these X-ray faint objects
also appear to be large, sometimes formally above the Eddington value,
based on the best black hole mass estimates. This also seems to be the
case for Mrk~231. Assuming the dynamically derived black hole mass of
1.7$^{4.0}_{-1.2} \times$ 10$^7$ M$_\odot$ based on stellar velocity
dispersions measured from near-infrared spectra (Dasyra et al.\ 2006),
the Eddington ratio of the AGN in Mrk 231 is 5.0$^{+11.3}_{-3.5}$.
The $\sim$5 $\times$ higher value of M$_{\rm BH}$ from Kawakatu et
al.\ (2007) would bring the Eddington ratio closer to unity; this mass
measurement is derived from the H$\beta$ line width and optical
continuum and is considered an upper limit because it was taken using a
3\arcsec\ aperture and is no doubt affected by continuum-light
contamination from the host. More recently, Leighly et al.\ (2014)
derived a black hole mass M$_{\rm BH}$ = 2.3 $\times$ 10$^8$ $M_\odot$
from the width of broad Pa$\alpha$ and the 1 $\mu$m luminosity density
(using the prescription of Landt et al.\ 2013); this results in an
Eddington ratio of $\sim$0.3. Sources with near or super-Eddington
accretion rates are expected to drive nuclear outflows. Mrk~231 fits
the mold given its FeLoBAL characteristics, indicative of a nuclear
outflow of up to 8000 km s$^{-1}$, and tentatively detected ultra-fast
X-ray wind of 20,000 km s$^{-1}$ (Feruglio et al.\ 2015).

Our new FUV data indicate that Mrk~231 shares at least two additional
properties with these fast accreting X-ray weak objects: the FUV
emission lines of Mrk~231 are weak relative to the continuum and
highly blueshifted. As listed in Table 1, the equivalent width of C~IV
measured in Mrk~231 is only 12 \AA. That is $\sim$3 $\times$ smaller
than the typical value of luminous SDSS quasars (30.0 $\pm$ 0.03 \AA;
e.g., Vanden Berk et al.\ 2001; Richards et al.\ 2011). Given the
well-known Baldwin effect (i.e. quasars of lower luminosities present
larger C~IV equivalent widths on average), the difference in EW(C~IV)
between Mrk~231 and quasars of comparable luminosities is actually
even larger. Similarly, the blueshift of $\sim$3000 km~s$^{-1}$ of the
Ly$\alpha$, C~IV, and \ciii\ lines in Mrk~231 falls well above the
distribution of velocities in radio-quiet and radio-loud quasars
(e.g., Fig.\ 2 of Richards et al.\ 2011). In contrast, C~IV equivalent
widths $\la$12 \AA\ and blueshifts $\ga$ 1000 km s$^{-1}$ are common
in PHL~1811 analogs (Leighly et al.\ 2007b; Wu et al.\ 2011) and in
WLQs (Luo et al.\ 2015). Moreover, Plotkin et al.\ (2015) have
recently shown that the low-ionization lines of H$\alpha$ and H$\beta$
in PHL-1811 analogs and WLQs are not exceptionally weak, in contrast
to their high-ionization lines like C~IV. Mrk~231 also shows very weak
to no high-ionization line emission in the optical (e.g., He II 4686
\AA, [O~III] 5007 \AA; e.g., Rupke et al.\ 2002), near-infrared (e.g,
   [Si~VI] 1.96 $\mu$mm; Leighly et al.\ 2014), or mid-infrared (e.g,
   [O~IV] 25.89 $\mu$m, [Ne~V] 14.32, 24.32 $\mu$m, [Ne~VI] 7.65
   $\mu$m; Armus et al.\ 2007), but relatively normal low-ionization
   optical line emission. Finally, PHL-1811 analogs and WLQs also
   present relatively strong optical and NUV Fe~II emission, as seen
   in Mrk~231. As discussed in Section 4.2., these similarities
   between Mrk~231 and ``wind-dominated'' PHL~1811 analogs and WLQs
   provide critical new clues on the inner structure of the AGN in
   Mrk~231.

\subsection{Physical Model -- Revisited}

In Figures 5 and 6 of Paper I, we presented two simple geometrical
models that could explain the main features observed in the 2011 COS
data, namely the peculiar Ly$\alpha$ emission, the flat and nearly
unpolarized FUV continuum, and the absence of broad absorption
features in the FUV despite the presence of a strong optical-NUV
FeLoBAL outflow. The main difference between the two pictures was the
geometry of the BALR. In the simplest picture, we assumed a spherical
geometry for the BALR, while a more physically motivated disk geometry
was assumed in the second case. In both pictures, the outflowing BALR
was assumed to act as both the partial screen for the continuum
emission and the source of the observed blueshifted Ly$\alpha$
emission.  In both cases, the source of the FUV continuum was assumed
to be an accretion disk on scale of $\sim$few $\times$10$^{15}$ cm
($\la$0.01 pc).

Here we refine these pictures taking into account the new constraints
from our Cycle 21 {\em HST} data and the similarities we found between
Mrk~231 and wind-dominated WLQs (Section 4.1).  Another important
piece of information that has come to our attention since the
publication of Paper I is the detection of strong radio flares in
Mrk~231 at 20 GHz (Reynolds et al.\ 2009; 2013) and, more recently, at
1 and 3 mm (Lindberg et al.\ 2016). Reynolds et al.\ (2009, 2013) have
argued that these flares have blazar-like characteristics and are
therefore associated with highly relativistic ejecta whose
self-absorbed synchrotron radio emission is enhanced by strong Doppler
boosting (Doppler factor $\delta$ $>>$ 1). This is surprising since
Mrk~231 is formally a radio-quiet quasar, and more precisely, a BALQSO
in which the radio jets are typically suppressed (although see Becker
et al.\ 1997).  Given a time variability brightness temperature of
$T_b$ = (12.4 $\pm$ 3.5) $\times$ 10$^{12}$ K and the need to avoid
the inverse Compton catastrophe (Marscher et al.\ 1979; Ghosh \&
Punsly 2007), Reynolds et al.\ (2009, 2013) constrained the jet angle
to be less than $\theta_{\rm max}$ = 25.6$^{+3.2}_{-2.2}$ from the
line of sight.  This implies a nearly face-on view of the inner
accretion disk, assuming the jet is directed along the rotation axis
of the inner disk. In this picture, the pc-scale double radio
structure detected in VLBA observations is made of a core component
that experiences flares and a steep spectrum secondary component
corresponding to a compact radio lobe that is associated with the
working surface between the confined jet and dense ISM. Warping of the
disk structure on larger scale may be responsible for the apparent
discrepancies in orientation of structures on scales $\ga$ 100 pc
(e.g., Carilli, Wrobel, \& Ulvestad 1998; Downes \& Solomon 1998;
Klockner, Baan, \& Garrett 2003; Davies et al.\ 2004).

This idea of pole-on BAL outflows is not unique to Mrk~231. Ghosh \&
Punsly (2007) have used the same time-variable $T_B$ argument as above
to argue that an inordinately large fraction of high-$T_b$ BAL QSOs
are LoBALQSOs and therefore fall in this category of pole-on
outflows. In a more recent paper, DiPompeo, Brotherton, \& De Breuck
(2012) have argued that there is a distribution of BAL outflow viewing
angles that includes near pole-on, based on radio spectral index (see
also Berrington et al.\ 2013 for another example of pole-on BALQSO).

This interpretation of the radio flares in Mrk~231 first seems at odds
with the rather edge-on view of the disk ($i \ga 45^\circ$) favored in
the disk scenario discussed in Paper I (as well as Gallagher et
al.\ 2002) to explain the absence of redshifted Ly$\alpha$ emission
(in this scenario, the far-side portion of the dust-free BALR, where
Ly$\alpha$ is emitted, was largely obscured by the dusty BALR on the
near side).  This constraint on the orientation can be relaxed if the
opening angle subtended by the outflowing dust-free material is
narrower than previously assumed. This is precisely what is expected
in a system like Mrk~231, which is experiencing near- or
super-Eddington accretion (Section 4.1). When $L/L_{\rm Edd} \ga$ 0.3,
photon trapping and advection become important and the accretion flow
becomes geometrically thick, i.e. characterized by narrow low-density
funnels along the rotation axis (e.g., Abramowicz et al.\ 1988; Ohsuga
et al.\ 2005; Jiang et al.\ 2014; Sadowski et al.\ 2013, 2014, 2015;
McKinney et al.\ 2014, 2015; Wang et al.\ 2014).  The strong
self-shadowing effects of these so-called ``slim disks'' lead to
strong anisotropy of the radiation field, creating ``polar radiation
cones'' and polar radiation-driven outflows. In systems with
spinning supermassive black holes, these sub-relativistic outflows
coexist with electromagnetic relativistic jets driven by the
Blandford-Znajek effect.

A sketch of the revised disk geometry for Mrk~231 is shown in Figure
8. We warn the readers that this sketch is just meant to be
  illustrative; it is not accurate in detail. Figure 8 is very
similar to that shown in Figure 6 of Paper I, except that the disk
structure is now geometrically thicker and viewed more face-on than in
Paper I. In this picture, the polar outflow is the source of clouds
for the BALR.  The BELR is comprised of two physically related
components, the extended accretion disk atmosphere and the polar wind
(e.g., Murray et al.\ 1995; Murray \& Chiang 1995, 1997; Proga et
al.\ 2000; Chelouche \& Netzer 2003; Proga \& Kallman 2004; Proga
2007; Proga, Ostriker, \& Kurosawa 2008; Richards et al.\ 2011;
Kruczek et al.\ 2011; Richards 2012; Kashi et al.\ 2013; Wang et
al.\ 2014). The driving mechanism(s) of the polar wind is
intentionally left unspecified: while the previously cited models
favor radiation pressure driving, magnetohydrodynamic (MHD) driving
cannot be ruled out (e.g., Fukumura et al.\ 2015 and references
therein).  The outflowing clouds and filaments labeled ``dust-free''
in this figure are directly exposed to the beamed X-ray + EUV
radiation field emitted in the twin low-density funnels of the slim
disk.  Clouds further from the rotation axis and labeled ``dusty'' are
less influenced by the polar radiation cone and have
proportionally smaller outflow velocities relative to their rotational
(virialized) velocities.  Finally, gas in the extended accretion disk
atmosphere and labeled ``disk BELR'' is shielded from (most of) the
EUV and soft X-ray radiation and shares the largely rotational
(virialized) motion of the slim disk material.

Figure 8 assumes that all of the EUV-FUV and soft X-ray radiation is
produced in the funnels of the slim disk, while the NUV continuum is
produced immediately outside of these funnels and the optical
continuum is produced further out in the slim disk. Also, no attempt
is made to take into account the (possibly important) effects of dust
sublimation near intense NUV sources. So, in reality, the division
between dust-free and dusty clouds may not be as simple, and the disk
BELR likely lies further from the center than the NUV continuum source
(to be consistent with results from reverberation mapping; e.g.,
Peterson et al.\ 2004). The outflowing dusty clouds correspond to the
FeLo BALR and are also the source of the observed blueshifted FUV line
emission, including Ly$\alpha$, \ciii\, C~II $\lambda$1334, and C~IV
$\lambda$1548 (the observed Ly$\alpha$ emission from these clouds must
be produced near the surfaces of these clouds to avoid destruction by
dust grains).  In contrast, the disk BELR is the source of most of the
optical broad-line emission. The small solid angle subtended by the
highly ionized gas exposed to the EUV and X-ray radiation emerging
from the narrow funnels of slim disks may explain the general deficit
of high-ionization emission and absorption lines in Mrk~231, PHL-1811
analogs and WLQs (Luo et al.\ 2015).

To explain the blazar-like radio flares of Mrk~231, a relativistic jet
must also be present along the rotation axis of the accretion disk,
and our line of sight to the disk must be close to that axis ($\la$
26$^\circ$; Reynolds et al.\ 2013). The compact radio lobe located
within $\sim$1 pc from the primary radio component is not shown in
Figure 8 for clarity. On the other hand, the weakness of the observed
FUV and X-ray continua and deduced EUV continuum (from the faint
high-ionization lines) strongly argues against a direct view of the
beamed ionizing radiation emerging from the twin funnels of the slim
disk (e.g., Costell\'o-Mor, Netzer, \& Kaspi 2016). This puts a lower
bound on the angle of our line of sight to the rotation axis of order
$\ga$10$^\circ$, although this number depends on the spin of the
supermassive black hole and also on the exact structure of the slim
disk, which is model-dependent (e.g., Jiang et al.\ 2014; Sadowski et
al.\ 2013, 2014, 2015; McKinney et al.\ 2014, 2015).

In this picture, the outflowing dusty BALR acts as a partial screen
for the FUV-NUV-optical continuum and BELR emission.  A covering
fraction of $\sim$90\% was estimated in Rupke et al.\ (2002) and Paper
I based on the non-zero flux in the core of the saturated boxcar
profile of Na~I~D measured from the high-resolution Keck
spectrum. Given the similar physical conditions leading to Na~I
(ionization potential of 5.1 eV) and Mg~I (7.6 eV) absorption, one can
infer that the NUV absorbers also partially cover the continuum source
(although the low spectral resolution of the STIS data does not allow
us to directly estimate the covering fraction in the NUV).  The
  profiles of the unblended absorption features (Mg~I, He~I, Ca~II H,
  and Ca II K) are distinctly narrower than those of the FUV emission
  lines. In this picture (Fig.\ 8), the widths of these absorption
  features reflect the range of velocities of the BALR clouds along
  our line of sight to the optical-NUV source, while the broader
  profiles of the FUV emission lines reflect the velocities of all
  BALR clouds in front of the disk, including the many clouds that are
  not directly along our line of sight to the optical-NUV source. The
  absence of redshifted Ly$\alpha$, \ciii\, and C~IV emission is due
  to obscuration of the disk BELR by the near-side dusty BALR
  clouds. The fact that the bulk of the BALR clouds cluster around
  velocities $\sim$ [--5500, --3500] km s$^{-1}$ suggests that the
  BALR screen is kinematically detached from the continuum source and
  does not extend over a broad range of distance from the disk.

The patchy BALR is opaque to the FUV but not to the optical-NUV
continuum, consistent with the smaller extinction and polarization in
the FUV than in the NUV reported in Paper I and Smith et al.\ (1995),
respectively.  The apparently smaller extinction towards the optical /
near-infrared continuum relative to the NUV continuum (Leighly et
al.\ 2014) could also be explained in this scenario if the source of
the optical / near-infrared continuum emission (e.g., emission from
the accretion disk and/or host galaxy) is more extended / less
affected by the dusty BALR than the NUV continuum region. This near
face-on view of the accretion disk is also consistent with the
relatively small column density recently derived from the {\em
  Chandra} + {\em NuSTAR} data ($N_H$ $\sim$ 1 $\times$ 10$^{23}$ 
  cm$^{-2}$; Teng et al.\ 2014), as it avoids the highly Compton
thick material in the slim disk. In this picture, about 10\% of
  the FUV emission makes it through the BALR, perhaps along the
  near-side funnel (line of sight \#1 in Figure 8).

The observed wild variations in the X-ray and radio (mm- and cm-wave)
bands but lack of NUV and FUV continuum and line variability over time
scales of up to 22 years provide additional constraints on the
structure and duty cycle of this AGN. While the radio variability is
attributed to strong Doppler boosting of self-absorbed synchrotron
emission from a large injection of relativistic particles (Reynolds et
al.\ 2009, 2013), the X-ray variations are best modeled by changes in
absorbing column densities and/or covering factor of the absorbing
screen in front of the disk funnel (e.g., Piconcelli et al.\ 2013;
Teng et al.\ 2014). The disk disruption / jet ejection model used to
explain the anti-correlation between radio and X-ray emission in some
jetted AGN (e.g., 3C~120; Lohfink et al.\ 2013) therefore may not
apply to Mrk~231. The stability of the FUV-NUV continuum emission and
NUV absorption troughs argue in favor of a thick and uniform BALR
screen made of numerous clouds and filaments (cf.\ McGraw et
al.\ 2015).

While a full analysis of the NUV BAL features (see Figures 2 --  3) is
not possible here given the poor spectral resolution of the
STIS data, some general statements can be made about the properties of
the BALR.  The BAL features of the ground-state resonant lines Mg II,
Mg I, and Na I are strong and relatively simple.  Given that Mg I and
Na I are easily ionized, their detections in Mrk~231 imply that there
is very high density and well-shielded gas consistent with our picture
of dust in the outflow. Fe II from various excited states (UV60, 61,
62, 63, and 78) are also clearly present in the STIS spectrum, as is
often the case in other FeLoBAL outflows (e.g., Hall et
al.\ 2002). Conclusions drawn from the detailed analyses of some of
these other FeLoBALs may thus also apply to the outflow in Mrk~231.

The detection of He~I$^*$ $\lambda$3889 requires log $N_H$
  (cm$^{-2}$) $\ga$ 22.5 and large ionization parameters in an {\em
    ionized} He~II / H~II absorbing region (e.g., Leighly et
  al.\ 2011; Hamann et al.\ 2016, in prep.). Large columns are needed
  because this line arises from the metastable 2s $^3$S state that is
  19.8 eV above ground state.  Fe~II, Ca~II, Mg~I, and Na~I require an
  additional separate column of warm neutral gas, nominally behind the
  ionized He~II / H~II zone. However, Ca~II and Na~I (5.1 eV
  corresponding to $\sim$ 2400 \AA) are not radiatively shielded
  behind the H~II -- H~I recombination front. Dust appears to be
  required to provide this shielding in at least the Na~I absorbing
  region. Hamann et al.\ (2016, in prep.)  discuss the physical
  conditions needed for Na~I and other low-ionization lines in the
  context of another outflow quasar with many FeLoBAL-like
  features. Those results suggest that the Na~I lines in Mrk~231
  require a neutral column with log $N_H$ (cm$^{-2}$) $\sim 22 -
  22.5$ and enough dust to provide extinction equivalent to $A_V \sim
  1-2$ mag.

The Fe~II UV60, 61, 62, 63, and 78 lines arise from an excited state
at $\sim$1.0 eV, which requires $n_H \ga 10^5 cm^{-3}$ to populate
these levels and produce the lines we observe in Mrk~231 (Hamann et
al.\ 2016, in prep.). This is not a strong constraint, but if we
assume that this limit applies also to the ionized gas producing
He~I$^*$ $\lambda$3889, then the combination of a minimum density and
minimum ionization parameter (log $U \ge -1$) lead to a distance $\la$
2 -- 20 pc between the absorber and the ionizing continuum source. For
this calculation, we used equation (A1) from Hamann et al.\ (2011)
that relates $n_H$, $U$, and $\nu L_\nu$(1450), the luminosity at 1450
\AA\ as seen by the BAL clouds, and assumed $\nu L_\nu$(1450) = 0.6 --
90 $\times$ 10$^{43}$ erg s$^{-1}$ corresponding to extinctions $A_V
\sim 0 - 1$ mag.\ to our line of sight for Galactic, LMC, or SMC
extinction curves. This distance falls slightly below the range of
distances (13 -- 230 pc) derived by Leighly et al.\ (2014) using
photoionization models to reproduce Ly$\alpha$, He I$^*$ 3889 and
10830, Ca II H and K, Na~I~D, and Fe~II. The mass outflow rate and kinetic
luminosity of the BAL outflow derived in that paper therefore need to
be proportionally scaled down to $\la$ 10 -- 100 $M_\odot$ yr$^{-1}$
and $\la$ 10$^{44 - 45}$ erg~s$^{-1}$ to account for the smaller
distance of the absorbers inferred from the detection of Fe~II from
excited states.

\subsection{Evidence against the Binary Black Hole Model of Yan et al.\ (2005)}

An alternative model has recently been proposed by Yan et al.\ (2015)
to explain the peculiar optical-to-UV SED of Mrk 231:\ continuum
emission from accretion flows onto a tight binary black hole, with a
semimajor axis of $\sim$590 AU and an orbital period $t_{\rm orb}$
$\sim$1.2 yrs. In their model, the optical continuum is produced by a
circumbinary disk, while the FUV continuum is produced by a mini-disk
surrounding the secondary black hole. The sharp drop-off observed at
2500 -- 4000 \AA\ is interpreted as a flux deficit due to a gap (or
hole) opened by the secondary black hole migrating within the
circumbinary disk. Given the morphological evidence that Mrk~231 is
indeed a recent merger (e.g., Hamilton \& Keel 1987; Hutchings \& Nef
1987; Surace et al.\ 1998; Veilleux et al.\ 2002, 2006), and the
dearth so far of black hole binaries in the cores of quasars (e.g.,
Popovi\'c 2012; Eracleous et al.\ 2012; Shen et al.\ 2013; Liu et
al.\ 2014; Runnoe et al.\ 2015), contrary to theoretical expectations
(Begelman et al.\ 1980; Yu 2002; Merritt \& Milosavljevi\'c 2005), the
binary black hole explanation for the peculiar SED of Mrk 231 is
potentially a very exciting new explanation of the Cycle 19 data, so
it deserves to be examined in detail taking into account the new
constraints from the Cycle 21 data.

As pointed out by Yan et al.\ (2015), numerical simulations of black
hole binaries (e.g., Miller \& Krolik 2013; Farris et al.\ 2014ab;
Roedig et al.\ 2014) show that the accretion onto the secondary black
hole is expected to vary on a timescale of the order of the black hole
binary orbital period, $t_{\rm orb}$ $\sim$ 1.2 years, or
less. However, as described here and in Paper I, the UV continuum
emission in Mrk 231 has been surprisingly constant over the last
several years. There is no evidence for any flux variations in the FUV
continuum emission between 2011 (Paper I) and 2014 (this paper), or
$\sim$3 $\times$ $t_{\rm orb}$, and in the NUV continuum emission
between 1992 (Smith et al.\ 1995) and 2014 (this paper), or $\sim$18
$\times$ $t_{\rm orb}$. This is not to say that the UV continuum of
Mrk~231 does not vary at all. In fact, there is some evidence that the
continuum level near 1300 \AA\ in the 1978 archival {\em International
  Ultraviolet Explorer} ({\em IUE}) spectra of Mrk 231 (published in
Hutchings \& Neff 1987) was $\sim$1.5 $\times$ higher than the
continuum level measured in our recent COS FUV data. But, the binary
black hole model has difficulties explaining the stability of the UV
emission in recent years.  It would imply a (unrealistically) steady
accretion stream of gas from the inner edge of the circumnuclear
binary to the mini-disk of the secondary black hole over timescales of
$\sim$3 -- 18 $t_{\rm orb}$.

A potentially fatal hurdle of the binary black hole model is the
stability of the Ly$\alpha$ profile over a period of $\sim$3 yrs
($\sim$3 $\times$ $t_{\rm orb}$). In this model, the mini-disk of the
secondary black hole is expected to dominate the line emission in the
FUV, and the large blueshift of Ly$\alpha$, \ciii, and C~IV would
therefore be due to the expected large line-of-sight velocity of the
secondary black hole as it orbits the primary black hole. However, if
this were the case, one would expect the centroid of the Ly$\alpha$
profile to shift in velocity from --3000 to at least +3000 km s$^{-1}$
over half an orbital period. This is firmly ruled out by the new
Cycle 21 data.

Another difficulty with the binary black hole model is the fact that
the absorption lines are only observed in the NUV and optical, but not
in the FUV. This means that the BALR must be present in the direction
of the optical/NUV continuum source, the circumbinary disk, but not in
the direction of the FUV continuum source, the mini-disk of the
secondary black hole.  This seems hard to explain without invoking a
contrived geometry of the BALR. Moreover, one would naively expect the
velocities of the FUV emission lines to be significantly different
from the velocities of the NUV/optical absorption lines, but this is
not the case (Figure 7). Recently, Leighly et al.\ (2016) has also
argued that the binary black hole model has difficulties explaining
the strong broad near-infrared recombination line emission relative to
the C~IV
% and C~III] 
line emission.

Finally, it is worth repeating that the spectropolarimetric study of
Smith et al.\ (1995) find that the polarization level peaks at a value
of $\sim$15\% around 3100 \AA, decreasing in the UV from $\sim$3100 to
1600 \AA, becoming essentially zero at the shortest wavelengths. The
polarization also monotonically decreases from $\sim$3100 to
$\sim$7600 \AA. This latter trend with wavelength cannot be explained
via polarization from electron scattering off of an asymmetric
structure (e.g., edge of circumbinary disk; Yan et al.\ 2015). Smith
et al.\ (1995) argue that this trend is due to the frequency-dependent
scattering cross-section of dust grains, while the trend in the UV has
been explained in the past as dilution by either
unobscured/unpolarized starlight (Smith et al.\ 1995; Leighly et
al.\ 2014) or AGN emission (Paper I and this paper).  In the binary
black hole scenario, these spectropolarimetric results would again
seem to imply a contrived geometry of the dust distribution with
respect to the black hole binary, where the small-disk FUV emission is
somehow unobscured/unpolarized while the NUV-optical emission from the
larger circumbinary disk is obscured/polarized.

\section{Summary}

We report the results of our analysis of new far- and near-ultraviolet
($\sim$1150 -- 3000 \AA) spectra of Mrk~231 obtained with {\em HST}
COS and STIS in Cycle 21 (2014), as well as a near-contemporaneous
optical spectrum obtained with the 4.3-meter DCT. These data are
compared with our Cycle 19 (2011) COS FUV ($\sim$1150 -- 1470 \AA)
spectra published in Paper I and archival NUV data. The main results
of this analysis are:

\begin{itemize}

\item The Ly$\alpha$ and H$\alpha$ emission profiles and optical
  and FUV continuum emission have not varied significantly over the
  past $\sim$3 years. The Ly$\alpha$ emission is faint, broad ($\ga$
  10,000 km s$^{-1}$ at the base), and highly blueshifted (centroid at
  $\sim$ --3000 km s$^{-1}$), while H$\alpha$ is strong and highly
  symmetric, spanning a range in velocity that is at least as large as
  that of Ly$\alpha$.  The FUV continuum is featureless (i.e. no
  obvious absorption features of stellar or AGN origin) and nearly flat. 

\item The broader wavelength coverage of the new data now reveals that
  the FUV \ciii\ $\lambda$1909 and C~IV $\lambda$1548 emission
  features have line profiles that resemble that of Ly$\alpha$, while
  the profiles of the NUV He~I, Mg~I, Mg~II, and Fe~II absorption
  features resemble that of the optical Na~I~D, He~I$^*$
  $\lambda$3889, and Ca~II H and K lines, exhibiting broad blueshifted
  troughs that overlap in velocity space with the FUV emission-line
  features. The sharp upturn of the NUV-optical continuum emission
  above $\sim$2400 \AA, noted in Paper I and other papers but based
  only on archival data taken at different epochs, is also seen in the
  new near-contemporaneous data. A comparison with the 1992 {\em HST}
  FOS data of Smith et al.\ (1995) shows no evidence for changes in
  the NUV continuum emission and absorption features over the past 22
  years. Multiple lines of evidence indicate that the NUV emission is
  unresolved on scale $\la$40 pc. The flux of the STIS and COS data
  are consistent with each other in the overlap region, implying a FUV
  continuum-emitting size $\la$170 pc and again favoring an AGN origin
  for the FUV emission rather than a stellar origin.

\item These results strengthen the conclusions from Paper I that the
  NUV -- FUV continuum emission is produced predominantly by the AGN
  rather than stellar processes. The FUV continuum emission is
  partially ($\sim$90\%) blocked but otherwise unaffected by the dusty
  FeLoBAL screen, while the NUV continuum emission is filtered by this
  screen. The observed FUV line emission appears to be produced in the
  outflowing BAL cloud system, while the Balmer lines arise primarily
  from the standard broad emission line region seen through the dusty
  FeLoBAL.

\item The BAL detections in ground and excited states of Fe~II and
  He~I as well as in Na I, Mg~I, Mg II and Ca II require densities
  $\ga$ 10$^5$ cm$^{-3}$, ionization parameters log $U$ $\ge -1.0$,
  ionized gas column densities log $N_H$ (cm$^{-2}$) $\ga 22.5$,
  neutral gas column densities log $N_H$ (cm$^{-2}$) $\sim$ 22
  -- 22.5, and enough dust to provide extinction equivalent to at
  least $A_V$ $\sim$1 -- 2 mag. The lower limits on the density and
  ionization parameter imply a distance for the FeLoBAL screen of
  $\la$ 2 -- 20 pc from the ionizing source. The inferred mass outflow
  rate and kinematic luminosity of the BAL outflow is $\la$ 10 -- 100
  $M_\odot$ yr$^{-1}$ and $\la$ 10$^{44 - 45}$ erg~s$^{-1}$,
  respectively.

\item We revise the geometrical disk model discussed in Paper I,
  taking into account the constraints from our new data and the recent
  detection of blazar-like radio flares in Mrk~231. We draw an analogy
  with weak-lined ``wind-dominated'' quasars which share many of the
  peculiar UV and X-ray spectroscopic properties of Mrk~231, and have
  similar high (often super-Eddington) accretion rates. We favor a
  picture where the inner accretion structure of Mrk~231 is
  geometrically thick with twin narrow funnels that drive (radiation-
  or MHD-driven) polar winds, coexistent with electromagnetic
  relativistic jets. Our view of this ``slim disk'' must be nearly
  face-on to explain the high-frequency radio data. We associate the
  dust-free outflowing clouds in this picture with the absorbing
  material in the X-rays. The outflowing dusty clouds correspond to
  the FeLoBAL screen seen in the NUV-optical and are also the source
  of the observed blueshifted FUV line emission.  In contrast,
  the extended accretion disk atmosphere is the source of most of the
  optical broad-line emission.

\item The results of our analysis of these new data and comparisons
  with the older data are inconsistent with the predictions of the
  tight sub-pc binary AGN scenario proposed by Yan et al.\ (2015). The
  lack of FUV emission variability imply unrealistically steady
  accretion stream of gas from the inner edge of the circumnuclear
  binary to the mini-disk of the secondary black hole.  One would also
  expect the orbital motion of the secondary component around the
  primary to produce large velocity shifts in the Ly$\alpha$ emission
  profile over the orbital period $t_{\rm orb}$ $\sim$1.2 yr, but
  these shifts are not observed. The near-zero polarization and
  absence of BALs in the FUV, while the optical/NUV spectrum shows
  strong BALs and substantial polarization, would imply a very
  contrived geometry of the dusty FeLoBAL screen. The good match in
  velocities of the optical/NUV absorption lines and FUV emission
  lines seems inconsistent with this picture, regardless of the exact
  geometry.

\end{itemize}

Our results suggest that Mrk~231 is the nearest example of
``wind-dominated'' high accretion rate weak-lined quasars.
Super-critical black hole accretion is potentially relevant to a broad
range of phenomena including ultraluminous X-ray sources (ULX), tidal
disruption events (TDEs), and hyper-accreting supermassive black holes
at high redshifts.  However, the fractions of PHL-1811 analogs and
WLQs among typical quasars are small, 1\% $-$ 2\% (Wu et
al.\ 2011). It is not clear why that is the case.  Figure 8 suggests
that orientation effects may play a role (e.g., a more edge-on view of
Mrk~231 might obscure the fast wind). Another possibility is that the
conditions needed to reach near- or super-Eddington accretion rates
are generally not present in local quasars, except perhaps in some
rare luminous gas-rich merger events as in Mrk~231 (although see Du et
al.\ 2016 for other exceptions). In this context, Mrk~231 offers a
unique local laboratory to study a phenomenon that might be
fundamental to our understanding of the creation of supermassive black
holes in the early universe.

%\clearpage

\acknowledgements S.V.\ acknowledges useful discussions regarding the
high-frequency radio data of Mrk~231 with Drs.\ B.\ Punsly and
C.\ Reynolds. We thank J.\ I.\ Capone and Dr.\ S.\ B.\ Cenko for
obtaining the 2015 DCT optical spectrum and reducing it,
respectively. We also thank the anonymous referee for useful
suggestions that have improved this paper. Support for this work was
provided to S.V.\ and M.M.\ by NASA through contract {\em HST}
GO-13460.001-A.  S.V.\ also acknowledges support from the National
Science Foundation through grant AST-1207785. This work made use of
the Discovery Channel Telescope at Lowell Observatory. Lowell is a
private, non-profit institution dedicated to astrophysical research
and public appreciation of astronomy and operates the DCT in
partnership with Boston University, the University of Maryland, the
University of Toledo and Northern Arizona University. It also made use
of NASA's Astrophysics Data System Abstract Service and the NASA/IPAC
Extragalactic Database (NED), which is operated by the Jet Propulsion
Laboratory, California Institute of Technology, under contract with
the National Aeronautics and Space Administration.

%\clearpage

\clearpage
\capstartfalse
\begin{deluxetable*}{lrrrrrr}
\tablecolumns{7}
%\tabletypesize{\scriptsize}
\tablecaption{Strongest UV and Optical Spectral Features in Mrk~231}
\tablewidth{0pt}
\tablehead{
\colhead{Line ID} & \colhead{Flux}
& \colhead{REW} & \colhead{$f_\lambda$} & \colhead{$V_{50}$} &
\colhead{$V_{84}$} & \colhead{$V_{98}$} \\
\colhead{(1)} & \colhead{(2)} & \colhead{(3)} & \colhead{(4)} &
\colhead{(5)} & \colhead{(6)} & \colhead{(7)}
}
\startdata
Ly$\alpha$ $\lambda$1215.67 & $+$78& $-$71& 1.1& $-$2771& $-$5179& $-$7071\\
Ly$\alpha$ $\lambda$1215.67 (COS 2011)& $+$74& $-$68& 1.1& $-$2739& $-$5167& $-$7039\\
C II $\lambda$1334.43 & $+$13& $-$11& 1.2& $-$3232& $-$4565& $-$6143\\
C IV $\lambda$1548.20 & $+$16& $-$12& 1.3& $-$3253& $-$4910& $-$6003\\
C III] $\lambda$1909.0 & $+$15& $-$14& 1.0& $-$3636& $-$5903& $-$7072\\
Fe II UV2 $\lambda$2382$^a$ & $-$21& $+$14& 1.5& $-$3810& $-$5865& $-$7121\\
% Fe II UV2 $\lambda$2382 (1992)& ...& ...& ...& ...& ...& ...\\
Fe II UV1 $\lambda$2599$^a$ & $-$59& $+$26& 2.2& $-$3907& $-$6155& $-$7407\\
% Fe II UV1 $\lambda$2599 (1992)& ...& ...& ...& ...& ...& ...\\
Fe II UV62/63 $\lambda$2739$^a$& $-$32& $+$12& 2.7& $-$3521& $-$5664& $-$6610\\
% Fe II (UV62/UV63) $\lambda$2739 (1992)& ...& ...& ...& ...& ...& ...\\
Mg II $\lambda\lambda$2796.3, 2803.4& $-$36& $+$12& 3.0& $-$4742& $-$5572& $-$6191\\
% Mg II $\lambda$2798.0 (1992)& ...& ...& ...& ...& ...& ...\\
Mg I $\lambda$2853.0& $-$23& $+$6.5& 3.6& $-$4620& $-$4997& $-$5317\\
% Mg I $\lambda$2853.0 (1992)& ...& ...& ...& ...& ...& ...\\
He I $\lambda$2945.11 & $-$19.8& $+$4.7& 4.2& $-$4143& $-$5018& $-$5814\\
He I$^*$ $\lambda$3888.65 (Keck 2001)& $-$58.9& $+$7.1& 8.3& $-$4565& $-$4967& $-$6068\\
Ca II K $\lambda$3933.66 (Keck 2001)& $-$56.1& $+$6.3& 8.9& $-$4498& $-$4775& $-$5059\\
Ca II H $\lambda$3968.47 (Keck 2001)& $-$41.6& $+$4.6& 9.0& $-$4491& $-$4699& $-$4985\\
Na I D $\lambda\lambda$5890, 5896 (DCT 2015)& $-$124& $+$20& 6.2& $-$4573& $-$5105& $-$6225\\
Na I D $\lambda\lambda$5890, 5896 (Keck 2001)& $-$126& $+$21& 6.0& $-$4451& $-$5005& $-$6639\\
% Na I D $\lambda$5889.95 (DCT 2015)& $-$124& $+$20& 6.2& $-$4573& $-$5105& $-$6225\\
% Na I D $\lambda$5889.95 (Keck 2001)& $-$126& $+$21& 6.0& $-$4451& $-$5005& $-$6639\\
H$\alpha$ $\lambda$6562.80 (DCT 2015) & $+$1579& $-$266& 5.9& $-$321& $-$2727& $-$6620\\
H$\alpha$ $\lambda$6562.80 (Keck 2001) & $+$1414& $-$255& 5.5& $-$33& $-$2325& $-$6087\\
\enddata
 
 \tablecomments{ All measurements refer to the Cycle 21 2014 COS and
   STIS data unless otherwise noted. Col.\ (1): Line ID and rest-frame
   wavelength of the feature in \AA\ used for the calculations of the
   velocities. In the cases of the multiplets (labeled as $^a$), the
   velocities are calculated with respect to the ground-state resonant
   transition.  Col.\ (2): Flux of the feature in 10$^{-15}$ erg
   s$^{-1}$ cm$^{-2}$ (positive value corresponds to emission).
   Col.\ (3): Rest-frame equivalent width of the feature in
   \AA\ (negative value corresponds to emission).  Col.\ (4):
   Continuum level at the position of the feature in 10$^{-15}$ erg
   s$^{-1}$ cm$^{-2}$ \AA$^{-1}$. Col.\ (5): 50-percential (centroid)
   velocity in km s$^{-1}$, a measure of the bulk velocity of the
   outflow, if present.  Col.\ (6): 84-percentile velocity in km
   s$^{-1}$, corresponding to 1 $\sigma$ blueward of the centroid
   velocity.  Col.\ (7): 98-percentile velocity in km s$^{-1}$,
   corresponding to 2 $\sigma$ blueward of the centroid velocity, a
   measure of the maximum velocity of the outflow, if present.  }

\end{deluxetable*}
\capstarttrue

\clearpage

\begin{figure*}
\epsscale{0.7}
\centering
\includegraphics[width=1.05\textwidth,angle=0]{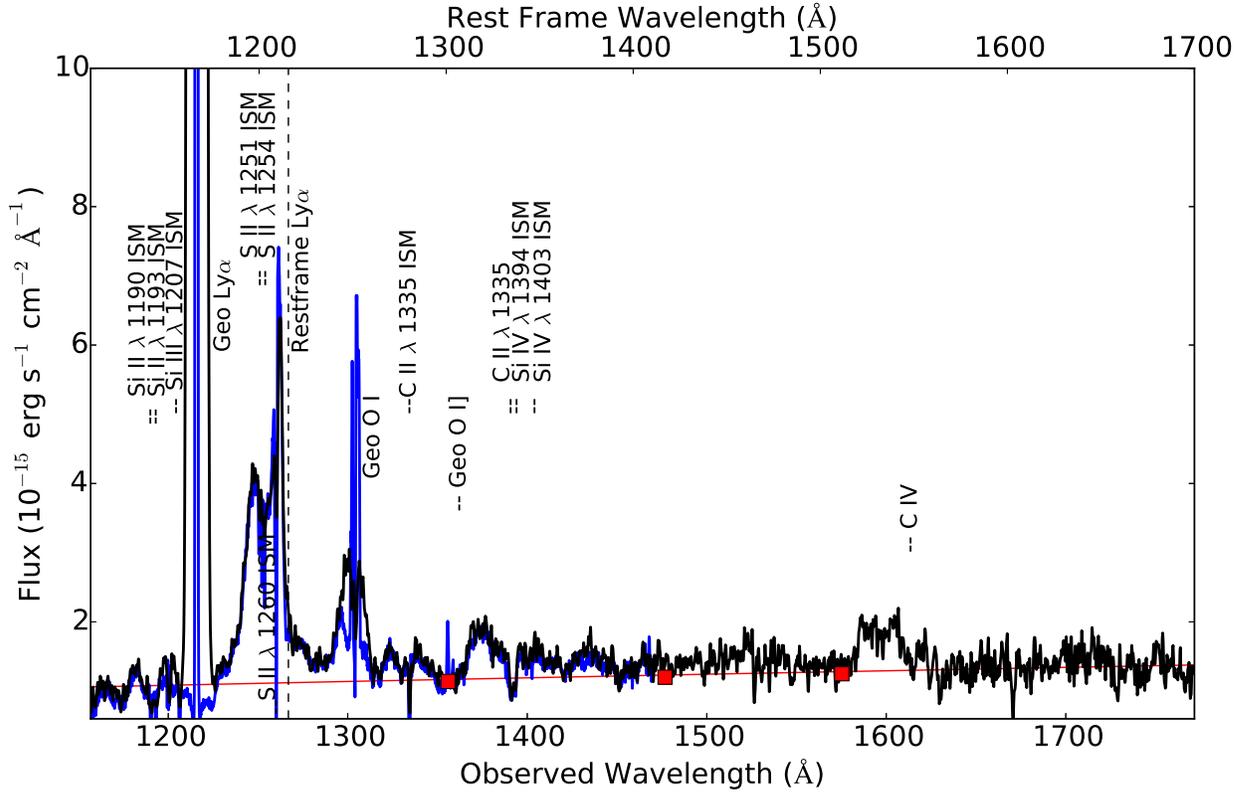}
\caption{{\em HST}-COS FUV spectra of Mrk~231, obtained in Cycle 19
  (blue) and Cycle 21 (black), binned to the same effective resolution
  to help with the comparison. These spectra have not been corrected
  for the small foreground Galactic extinction ($A_V \sim 0.03$ mag.).
  These spectra, taken $\sim$3 years apart, are remarkably
  similar. They are both dominated by broad, highly blueshifted
  Ly$\alpha$ emission (the vertical dashed line indicates the
  rest-frame position of Ly$\alpha$). Broad and highly blueshifted
  C~IV emission is also detected in the Cycle 21 spectrum. The FUV
  continuum emission is nearly featureless and only slowly declining
  at shorter wavelengths, consistent with $f_\lambda \propto
  \lambda^{0.7}$ (shown as a red line in the lower spectrum). It is
  dominated by the AGN and only slightly affected by dust reddening
  ($A_V \sim 0.5$ mag.; Paper I). }
\end{figure*}

\begin{figure*}
\centering
\includegraphics[width=0.8\textwidth,angle=0]{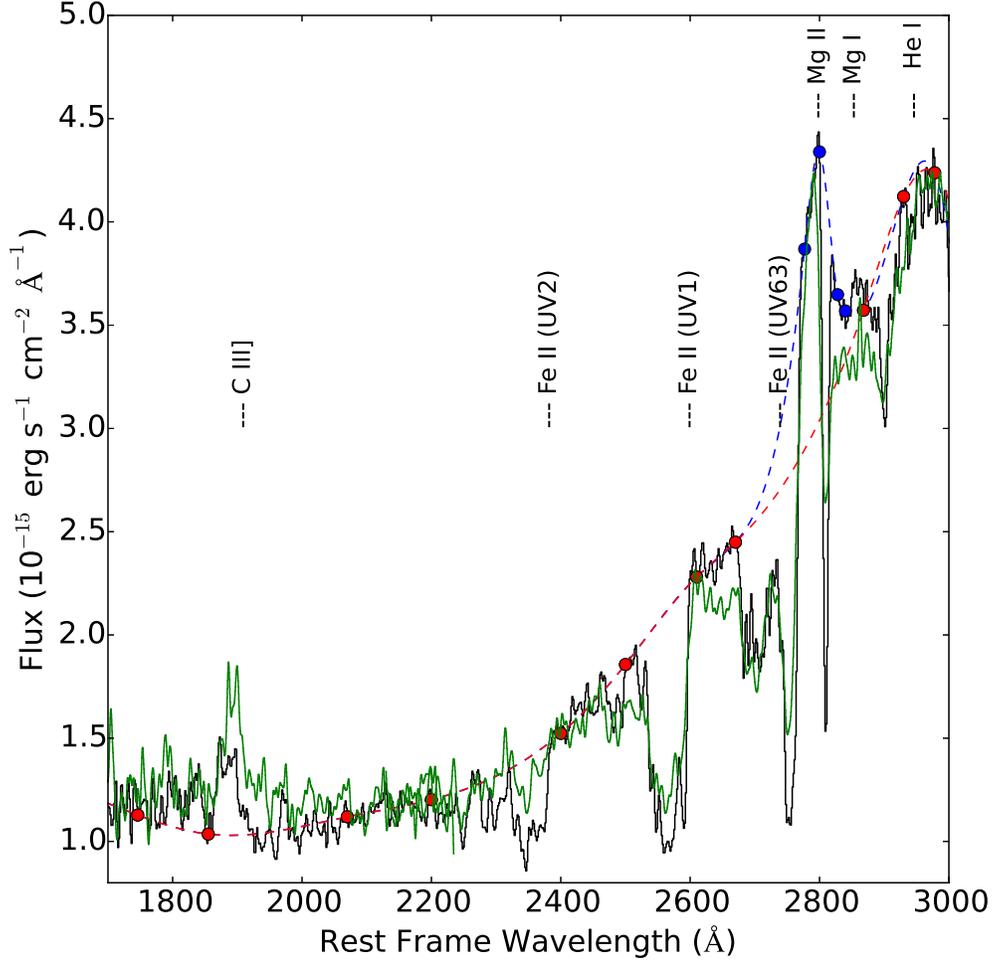}
\caption{{\em HST}-STIS NUV spectrum of Mrk~231 obtained in Cycle
  21. The STIS spectrum (shown in black) was lightly smoothed with a
  Gaussian kernel with a sigma (width) of 2 spectral pixels to help
  show the fainter features. The pre-COSTAR FOS spectra from Smith et
  al.\ (1995) are shown in green for comparison. The FOS G190H and
  G270H spectra were heavily smoothed and scaled to account for light
  losses due to pre-COSTAR aberrations and reproduce the spectrum
  shown in Figure 1 of Smith et al. The NUV spectrum of Mrk~231 is
  dominated by strong broad and blueshifted absorption-line features
  from He~I, Mg~I, Mg~II, and the Fe II UV1, UV2, and UV62/63
  multiplet transitions (the vertical dashed marks indicate the
  expected rest-frame positions of the resonant lines in these
  features). Broad and highly blueshifted C~III] emission is also
detected. The blue and red dashed curves represent the
pseudo-continuum fits that were used to determine the properties of
the broad absorption and emission features, respectively (see Table
1). }
\end{figure*}

\begin{figure*}
\includegraphics[width=1.0\textwidth,angle=0]{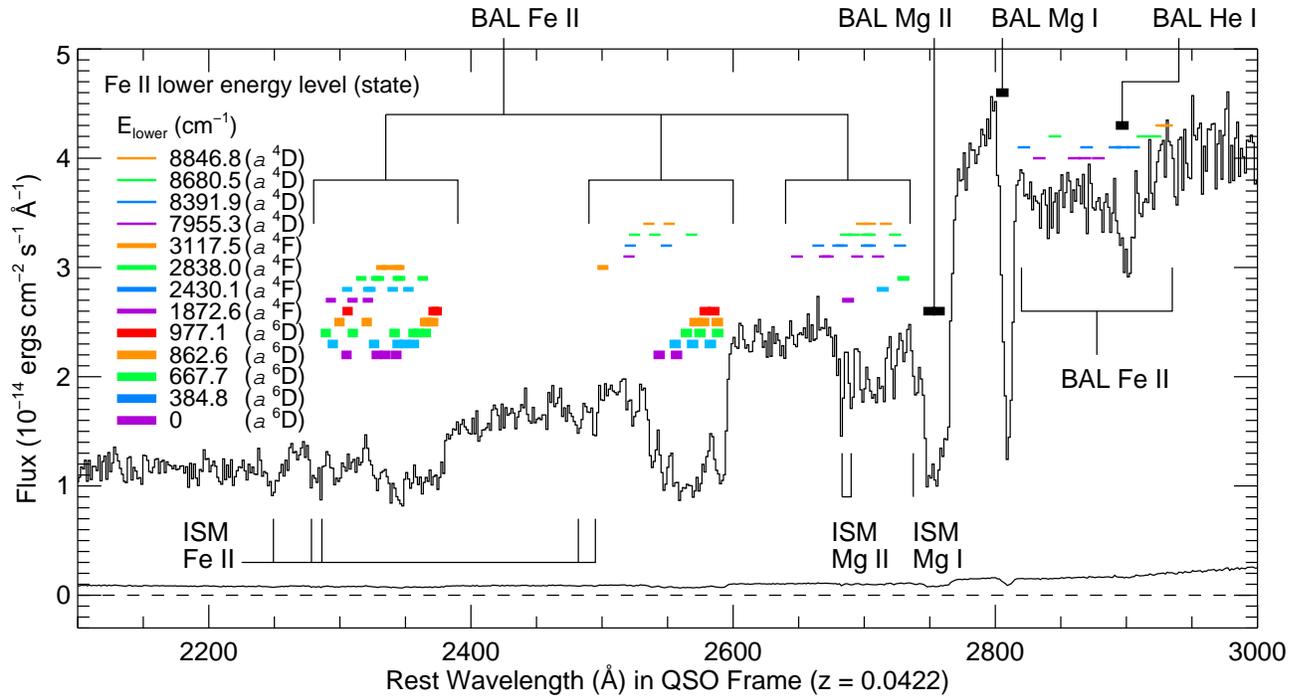}
\caption{The Cycle 21 STIS NUV spectrum of Mrk~231 is labeled to
  indicate the many transitions contributing to the Fe II
  multiplets. The thickness of each line indicates the lower state of
  each transition while the colors indicate the lower fine-structure
  energy level of each transition. The thickest lines are transitions
  arising from the a $^6$D ground state, intermediate-thickness lines
  arise from the metastable a $^4$F lower state, and the thinnest
  lines arise from the higher energy metastable a $^4$D state.  Also
  labeled on this figure are the positions of the BAL features from
  He~I, Mg~I, and Mg~II, and the narrow features produced by the
  Galactic ISM.}
\end{figure*}

\begin{figure*}
\epsscale{0.5}
\centering
\includegraphics[width=1.05\textwidth,angle=0]{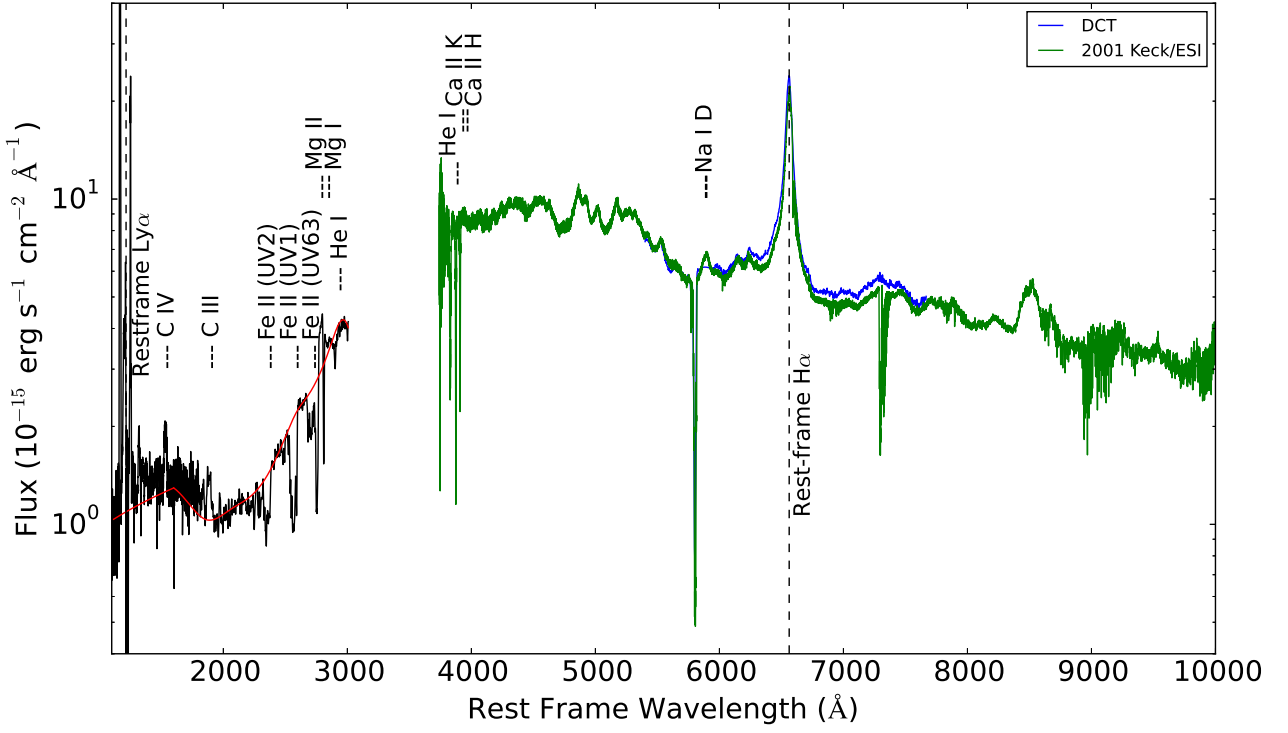}
\caption{ The UV -- optical spectral energy distribution (SED) of
  Mrk~231 in log~$f_\lambda$ {\em versus} log~$\lambda$(rest-frame)
  units.  This SED combines the new Cycle 21 COS and STIS data with
  the 2015 DCT (blue) and 2001 Keck/ESI (green) optical spectra.  Note
  the dramatic upturn in flux above $\sim$2200 \AA. The vertical dash
  lines indicate the expected rest-frame positions of the main
  spectral features. The red curve is the same as in Figure 1: it
  represents the pseudo-continuum used to measure the properties of
  the various FUV -- NUV absorption features (Table
  1). The discrepancy in flux between the DCT and Keck spectra at 6000
  -- 8000 \AA\ reflects errors in the flux calibration of the DCT
  data (which were acquired under non-photometric conditions).}
\end{figure*}

\begin{figure*}
\epsscale{1.0} \centering
\includegraphics[width=0.75\textwidth,angle=0]{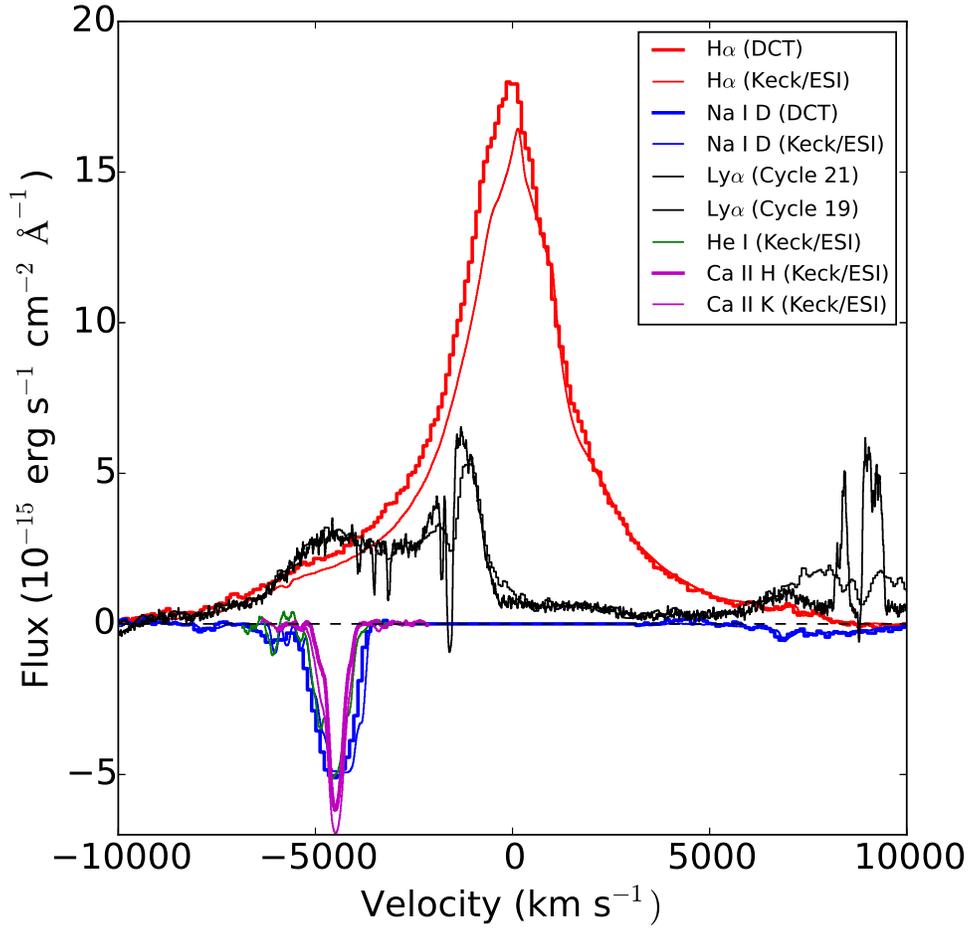}
\caption{Comparison of the continuum-subtracted H$\alpha$ emission
  (red), Ly$\alpha$ emission (black), Na I~D absorption (blue), He
  I$^*$ $\lambda$3889 absorption (green), Ca II H $\lambda$3968 (thin
  pink), and Ca II K $\lambda$3934 absorption (thick pink). All
  spectral features are on the same velocity and absolute flux
  scales. In all cases, the histograms represent the most recent {\em
    HST} and DCT spectra while the smooth curves are from the old Keck
  optical data (H$\alpha$, Na~I~D, He~I, and Ca~II H and K) or the
  Cycle 19 {\em HST}/COS data (Ly$\alpha$). Note that the
    Ly$\alpha$ profile at $\sim$ 5000 -- 10,000 km s$^{-1}$ is
    severely affected by geocoronal O~I residuals and is thus
    unreliable. The H$\alpha$ profiles in both the DCT and Keck data
  were interpolated at $\sim$ $+$1000 -- 2000 km s$^{-1}$ to remove
  the telluric O$_2$ $\lambda$6850 band. The slight discrepancy in the
  H$\alpha$ profile between the DCT and Keck data reflects errors in
  the flux calibration of the DCT spectrum (which was acquired under
  non-photometric conditions). As noted in Paper I, Ly$\alpha$ shares
  a stronger kinematic resemblance with the broad and blueshifed
  Na~I~D, He~I, and Ca~II absorption features than with H$\alpha$
  emission. See text and Table 1 for more detail.}
\end{figure*}

\begin{figure*}
\epsscale{1.0}
\centering
\includegraphics[width=0.65\textwidth,angle=0]{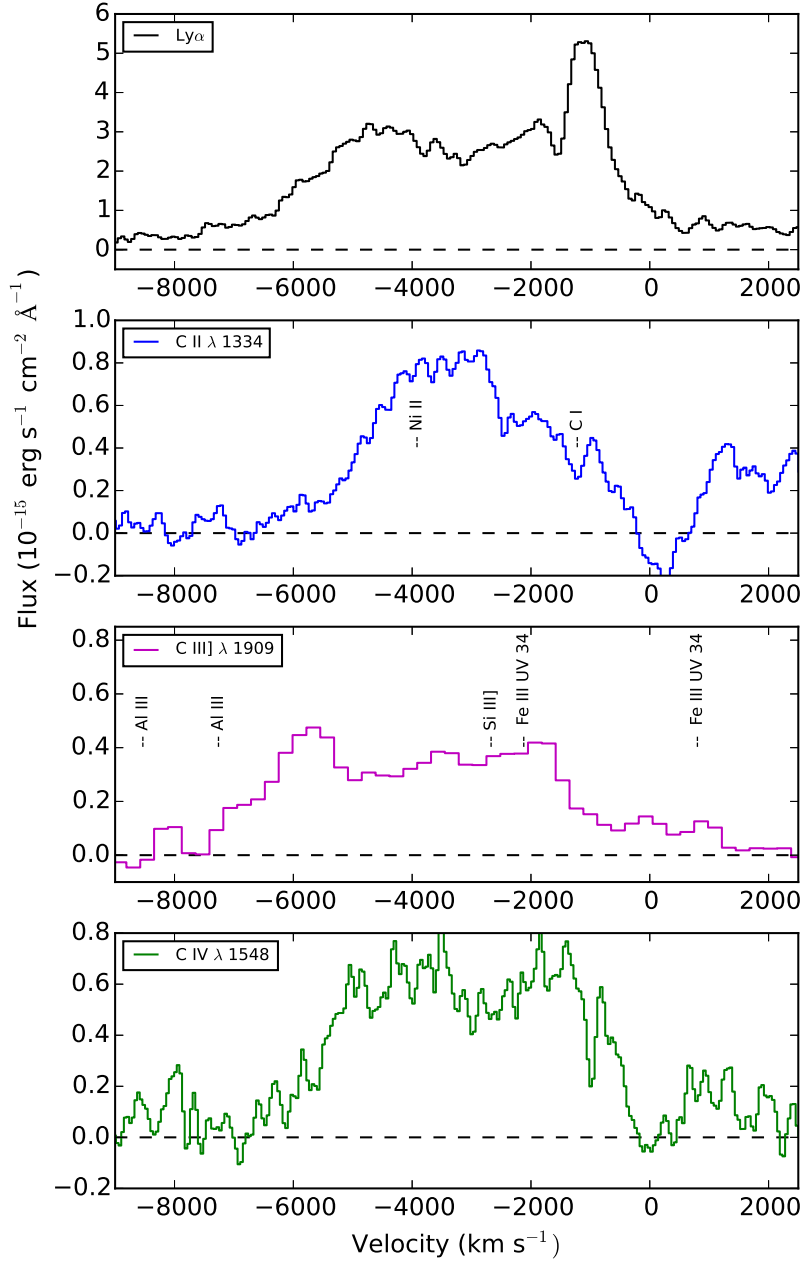}
\caption{Comparison of the continuum-subtracted FUV line emission from
  Ly$\alpha$ (black), C~II $\lambda$1334 (blue), \ciii\ $\lambda$1909
  (red), and C~IV $\lambda$1548 (green). All four features are on the
  same velocity scale but the vertical scale has been
  adjusted to better show the full range of absolute flux. The line labels
  in the panel showing \ciii\ indicate the $v = 0$ \kms\ positions of
  potentially contaminating features. See text and Table 1 for more detail.}
\end{figure*}

\begin{figure*}
\epsscale{1.0}
\centering
\includegraphics[width=1.0\textwidth,angle=0]{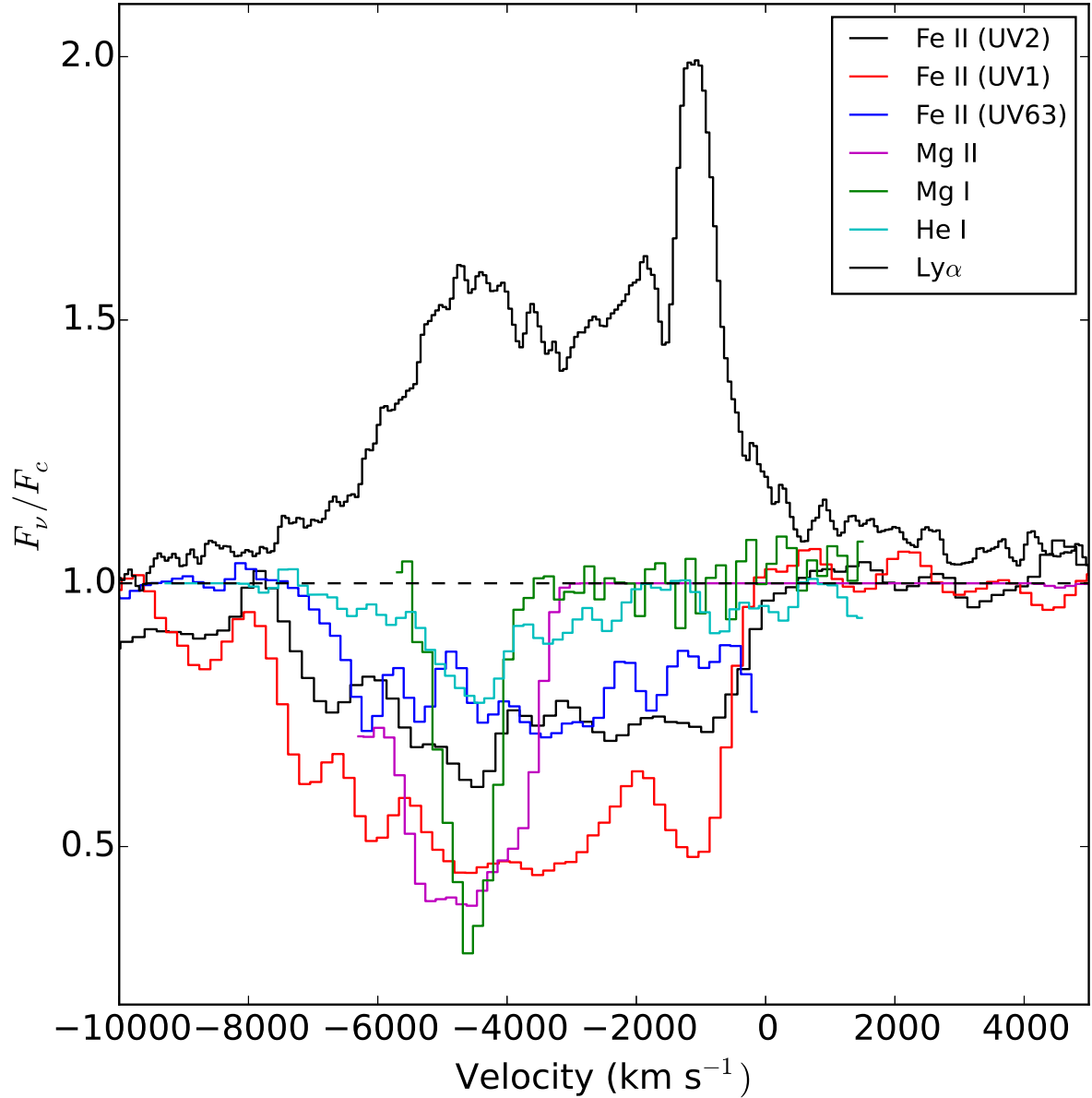}
\caption{Comparison of the continuum-normalized NUV line absorption
  from Fe II UV2 (black), Fe II UV1 (red), Fe II UV62/63 (blue), Mg II
  $\lambda\lambda$2796, 2803 (pink), Mg I $\lambda$2853 (green),
  and He~I $\lambda$2945 (turquoise) {\em versus} the Ly$\alpha$
  emission profile (black; the peak intensity of Ly$\alpha$ was
  arbitrarily normalized to unity). All of the features are on the
  same velocity scale. The presence of UV62/63 indicates strong
  absorption out of level $\sim$1 eV above ground and, therefore, all
  of the fine structure levels in UV1 and UV2 are probably also
  contributing absorption, not just the one resonance line. This
  accounts for much of the velocity extent of the Fe~II troughs. He I
  $\lambda$2945 also arises from an excited (metastable) state. See
  text and Table 1 for more detail.}
\end{figure*}

\begin{figure*}
\epsscale{1.0}
\centering
\includegraphics[width=1.0\textwidth,angle=0]{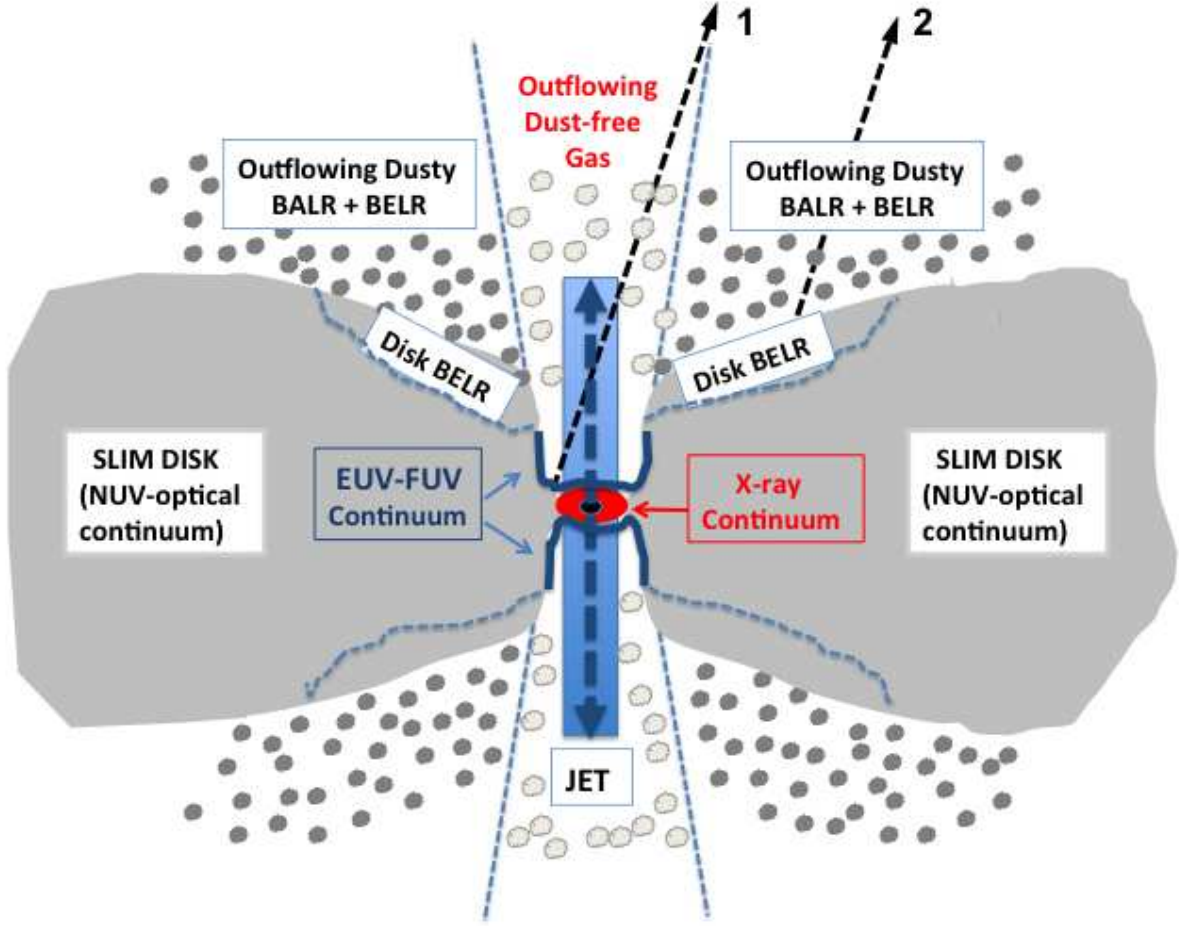}
\caption{ Revised geometric disk model of the central $\sim$10 pc
  region of Mrk~231 (not drawn to scale; we warn the readers that
    this sketch is just meant to be illustrative; it is not accurate
    in detail). The accretion flow is geometrically thick: it traces
  a ``slim disk'' characterized by twin narrow low-density
  funnels. Our line of sight, shown as two black dashed lines, is
  constrained to lie $\sim$10 -- 26$^\circ$ from the rotation axis
  based on the high-frequency radio and UV data --- our line of sight
  lies outside of the conical region defined by the near-side funnel
  but is close enough to the radio jet axis for relativistic boosting
  to be important. The outflowing dust-free material is highly ionized
  and not detected in the UV-optical spectrum of Mrk~231, but may
  contribute to the absorbing column measured in the X-rays. The
  outflowing dusty clouds act as the low-ionization BALR seen in the
  NUV-optical and are also responsible for the broad and blueshifted
  Ly$\alpha$, \ciii, and C~IV line emission. The BALR is opaque to the
  underlying FUV continuum emission but lets $\sim$10\% of this
    emission through, perhaps along the near-side funnel (e.g., line
    of sight \#1 has a direct peek at the FUV source). The line
  emission from the disk BELR is filtered through the dusty BALR and
  dominates the optical line emission (e.g., H$\alpha$). For clarity
  the pc-scale double radio structure detected in the VLBA
  observations is not shown in this image. See Section 4.2 for more
  detail. }
\end{figure*}

\clearpage

\end{document}